\documentclass[aps,prl,preprintnumbers,showpacs,nofootinbib]{revtex4}
\usepackage[english]{babel}
\usepackage{graphicx}
\usepackage{subfig}
\usepackage{dcolumn}
\usepackage{bm}
\begin{document}
\newcommand{\newc}{\newcommand}
\newc{\ra}{\rightarrow}
\newc{\lra}{\leftrightarrow}
\newc{\beq}{\begin{equation}}
\newc{\eeq}{\end{equation}}
\newc{\barr}{\begin{eqnarray}}
\newc{\earr}{\end{eqnarray}}
\def \lta {\mathrel{\vcenter
     {\hbox{$<$}\nointerlineskip\hbox{$\sim$}}}}
\def \gta {\mathrel{\vcenter
     {\hbox{$>$}\nointerlineskip\hbox{$\sim$}}}}
\def\vbf{\mbox{\boldmath $\upsilon$}}
\def\barr{\begin{eqnarray}}
\def\earr{\end{eqnarray}}
\def\g{\gamma}
\newcommand{\dphi}{\delta \phi}
\newcommand{\bupsilon}{\mbox{\boldmath \upsilon}}
\newcommand{\at}{\tilde{\alpha}}
\newcommand{\pt}{\tilde{p}}
\newcommand{\Ut}{\tilde{U}}
\newcommand{\rhb}{\bar{\rho}}
\newcommand{\pb}{\bar{p}}
\newcommand{\pbb}{\bar{\rm p}}
\newcommand{\kt}{\tilde{k}}
\newcommand{\kb}{\bar{k}}
\newcommand{\wt}{\tilde{w}}
\title{Transition Operators Entering  Neutrinoles Double Electron Capture to Excited Nuclear States}
\author{ J. D. Vergados$^{1,2}$ \thanks{Vergados@cc.uoi.gr}\footnote{Permanent address: Theoretical Physics Division, University
of Ioannina, Ioannina, Gr 451 10, Greece}}
\affiliation{$1$ Max Planck f${\ddot u}$r Kernphysik, Saupfercheckweg 1, 69120 Heidelberg,Germany}

\affiliation{$2$ CERN theory Division, Geneva, Switzerland}
\date{\today}
\begin{abstract}
We construct the effective transition operators relevant for neutrinoless double electron capture leading to final nuclear states different than $0^{+}$. From the structure of these operators we see that, if such a process is observed experimentally, it will be very helpful in singling out the very important light neutrino mass contribution from the other lepton violating mechanisms.
\end{abstract}
\maketitle
\section{Introduction}
		The nuclear double beta decay is  process considered very long time ago\cite{GOEPMAY}. It can occur whenever the ordinary (single) beta
		decay is forbidden due to energy conservation or greatly suppressed due to
		angular momentum mismatch. It could proceed with the emission of two neutrinos or in a neutrinoless mode. The 	$2\nu\beta\beta-$decay is just an example of a second order weak interaction. The exotic neutrinoless double beta decay 
		($0\nu\beta\beta-$decay) is the most interesting since it violates lepton number
		by two units.  It was first considered by Furry 
		\cite {Fur39} more than  half a century ago as soon it was realized that the
		neutrino might be a Majorana particle.
	To-day, seventy years later, $0\nu\beta\beta$-decay:
                \beq
                 (A,Z)\rightarrow~(A,Z+2)+e^-+e^-,\quad(0\nu~\beta\beta\mbox{  decay}),
                \label{eq:a}   
		\eeq
		$\Delta= Q $,
		 continues to be one of the most interesting processes, since:
		 \begin{itemize}
		 \item It will establish whether there exist lepton violating processes in the Universe. 
		 \item It is perhaps the only process, which can decide whether the neutrinos are Majorana particles, i.e. the particle coincides with its charge conjugate, or Dirac particles, if it does not. 
		 \item It will  aid in settling the outstanding question of the absolute scale of the neutrino masses, especially if this scale turns out to be very small. Neutrino oscillation experiments can only determine the mass squared differences.
		 \end{itemize}
		 For  recent reviews see \cite{JDV02,AEE08,Ejiri00,RODEJ11}.
		 
		  If this process is allowed other
                 related processes in which the charge of the nucleus
                 is decreased by two units may also occur, if they happen to be allowed by
                 energy and angular momentum conservation laws, e.g.
                 \beq
                 (A,Z)\rightarrow~(A,Z-2)+e^++e^+~~~~(0\nu\mbox{ positron emission}),
                \label{eq:b}   
		\eeq
		$\Delta=Q-4 m_e c^2$,
		 \beq
                 (A,Z)+e^-~\rightarrow~(A,Z-2)+e^+~~(0\nu~\mbox{ electron positron
                 conversion}),
                \label{eq:c}   
		\eeq
$\Delta=Q-2 m_e c^2$,
		 \beq
                 (A,Z)+e^-+e^-\rightarrow~(A,Z-2)+ x \mbox{rays }(0\nu\mbox{ double
                 electron capture }),
                \label{eq:d}   
		\eeq
      $ \Delta=Q$, where $\Delta$ is the available energy.
      
      The last process is always possible, whenever
                (\ref{eq:c}) is.
                 It has been first considered long time ago \cite{Ver83,BeRuJar83} as a two step process: In the first step the
                 two neutral atoms, (A,Z) and the excited (A,Z-2), get admixed
                 via the lepton number violating interaction .
                  In the second step the (A,Z-2) atom de-excites
                 emitting two hard x-rays and the nucleus, if it is found in
                 an excited state, de-excites emitting $\gamma$-rays. The life time expected was very long, since the above mixing amplitude was tiny compared to the energy difference of the two atoms involved. It  has recently, however, been gaining in importance \cite{SimKriv09,EliNovPC,DEC11,BELLI11} after ion Penning traps \cite{BLAUM06} made it possible to accurately  determine the $Q$ values, which gave rise to the the presence of resonances, which in turn could lead to an increase of the width my many orders of magnitude\cite{SimKriv09}. 
                Decays to excited states are in some cases possible
                and provide additional experimental information, e.g  $\gamma$ rays following
                their de-excitation.

Furthermore neutrinoless double electron capture is of experimental interest, since one expects the reaction to be essentially free of the two neutrino background, which plagues ordinary neutrinoless double beta decay.
 It has, also recently been realized \cite{EliNovPC} that the resonance condition can be fulfilled for transitions to excited states. Thus, e.g, in the case of $^{156}$Dy$\rightarrow ^{156}$Gd, such excited states can be\footnote{After completion of this work, two papers have been published \cite{KSFF11,EliNov}, which show that the resonance condition can be met for such excited states} $0^-,1^-,0^+,1^+$ and $2^+$.

Before proceeding further in examining neutrinoless double electron capture, we note the following:
\begin{itemize}
\item Transitions to  $0^-,1^-$  states
have not been considered previously in standard double beta decay, since they are energetically forbidden.
Such transitions can naturally occur, however, in the presence of right handed currents, arising by interference between the left handed and right handed currents, i.e. of the type $j_L \times j_R$, with an  amplitude is essentially independent of the neutrino mass. We will show that this process can occur in double electron capture, even though the electrons are bound.
\item Transitions to $0^+$ states, both the ground states and excited states, can occur both in the mass mechanism and the $j_L \times j_R$. The latter contribution in ordinary double beta decay occurs since the produced electrons have sufficiently high momenta. In double electron capture instead of the electron  momentum one can exploit the nuclear recoil term.
\item transitions to $1^+$ and $2^+$ can occur only, if the electron momentum is sufficiently high, as in neutrinoless double beta decay, or by retaining the nuclear recoil term in the hadronic current. In double electron capture this can naturally  happen, if one of the electrons is captured from a p-orbit.
\end{itemize}
 In the present paper we will derive the effective transition operators relevant for neutrinoless double electron capture, emphasizing the mode which is independent of the neutrino mass. The neutrino mass contribution has already previously examined, see e.g. the recent work and references therein \cite{SimKriv09}.
\section{Nuclear Transition Operators Leading to Negative Parity States}
To leading order such transitions cannot occur in neutrino mass mechanism. In the $j_L \times j_R$ case, however, the chiralities involved are such that in the intermediate neutrino propagator one picks the momentum, not the mass. The neutrino momentum leads to the following:
\begin{itemize}
\item  The time component of the momentum leads to an average energy difference $E_{n,1}-E_{n,2}$, where $E_{n,i}=\prec E_n\succ-M_i+E_{e}(i)$, with $ E_{e}(i)$ the electron energy, $i=1,2$.  This can lead to positive parity transitions in double beta decay, but it is not the dominant contribution.
\item the space part of the momentum leads to an effective operator which proportional to 
${\hat r}\frac{d}{dr}\frac{1}{r}$ with ${\bf r}={\bf r}_1-{\bf r}_2$, $r=|{\bf r}_1-{\bf r}_2|$. This naturally leads to negative parity transitions. 
\end{itemize}
 Positive parity transitions can, of course,  occur in neutrinoless double beta decay via this mechanism by retaining in the dipole term in the expansion of the essentially distorted plane wave function  of the two electrons, which leads to the term $i \left ( {\bf p}_1 .{\bf r}_1 +{\bf p}_2 . {\bf r}_2 \right ){\hat r}_{12}\frac{d}{dr}\frac{1}{r}$.  This possibility does not exist in neutrinoless double electron capture. We will, however, examine alternate possibilities in the next section.

The effective lepton violating transition operator, within  the closure approximation over the intermediate states, which holds for neutrinoless double beta decay and is expected to be even better in neutrinoless double electron capture, can be cast in the form:
\beq
{\cal M}=\frac{1}{\sqrt{2}} \left ( \frac{G_F}{\sqrt{2}}\right )^2 2 \frac{2i}{8 \pi R^2_0}u^R_{\mu \nu k}
  \sum_{n,m}\phi_{n_1,0}(r_n)\phi_{n_2,0}(r_m)J^{\mu}_R(n) J^{\nu}_L(m) {\hat r}^k_{n,m} f(x_{n,m})+L\leftrightarrow R
\eeq
with $\phi_{n_1,0}(r_n)\phi_{n_2,0}(r_m)$ ($r_n=r_1,\quad r_m=r_2$) the bound $s$-electron wave functions and
\beq
u^R_{\mu \nu k}=\bar{u}(p_1)\gamma_{\mu}\gamma_{k}\gamma_{\nu}(1-\gamma_5)u^c(p_2),\quad u^L_{\mu \nu k}=\bar{u}(p_1)\gamma_{\mu}\gamma_{k}\gamma_{\nu}(1+\gamma_5)u^c(p_2)
\eeq
\beq
J^{\mu}_L(n)=\tau_-(n)(g_V,-g_A \sigma_n),\quad J^{\nu}_R(m)=\tau_-(m)(g_V,+g_A \sigma_m)
\eeq
\beq
f(x_{n,m})=\frac{d}{d x} \frac{1}{x} J(\delta_0),\quad x=x(n,m)=|{\bf r}_n-{\bf r}_m|/R_0
\eeq
\beq
J(\delta_0)=\frac{2}{\pi}\int_0^{\infty}\frac{\sin{y}}{y+\delta_0} dy,\quad \delta_0=(\prec E_n\succ-M_i)|{\bf r}_n-{\bf r}_m|
\eeq
in double electron capture:
$$\delta_0\approx0,\quad J(\delta_0)\approx 1 \mbox{ and } f(x)\approx - \frac{1}{x^2}$$

For non relativistic electrons one  finds that the effective transition operator can be cast in the form:
\begin{enumerate}
\item An operator associated with the time structure of the leptonic current:
\beq
\ell_0={\bar u}(p_1)\gamma_0\gamma_5u^c(p_2),
\eeq
which is
\beq
\tilde{\Omega}_0=\frac{G_F^2}{2 \sqrt{2}}\frac{i}{\pi R^2_0} 4 g^2_A\sum_{n,m}\phi_{n_1,0}(r_n)\phi_{n_2,0}(r_m)\tau_-(n)\tau_-(m) 
{\hat r}_{n m}.\omega_0(n,m) f(x_{n,m}),\quad \omega_0(n,m)=i(\sigma_n\times\sigma_m)
\label{omega0}
\eeq
which can cause $ 0^+\rightarrow 0^-$ transitions.
\item An operator associated with the space structure of the leptonic current:
\beq
\vec{\ell}={\bar u}(p_1)\vec{\gamma} u^c(p_2)
\eeq
that is
\barr
\tilde{\Omega}_1&=&\frac{G_F^2}{2 \sqrt{2}}\frac{i}{\pi R^2_0}  g_V g_A\nonumber\\
& &\sum_{n,m}\phi_{n_1,0}(r_n)\phi_{n_2,0}(r_m) 2\tau_-(n)\tau_-(m) 
\left [{\hat r}_{n m}\times \omega_1(n,m) \right ] f(x_{n,m}),\quad \omega_1(n,m)=(\sigma_n-\sigma_m)\nonumber\\
\label{omega1}
\earr
\end{enumerate}
Since the $0s$ electron wave functions vary slowly inside the nucleus one can replace them by there average value and write:
\beq
\tilde{\Omega}_i=\Lambda_{\mbox{\tiny{eff}}} \prec {(0 s)^2}{a_B ^3}\succ \Omega_i, \quad \Lambda_{\mbox{\tiny{eff}}}=\quad \frac{G_F^2}{2 \sqrt{2}}\frac{i}{\pi R^2_0 a_B^3}
\eeq
where the $ \Lambda_{\mbox{\tiny{eff}}}$ has dimensions of energy and the average is over the radial part alone (the integral over the angles is unity). The dimensionless effective transition operators now become:
\beq
\Omega_0=4 g^2_A\sum_{n,m}\tau_-(n)\tau_-(m) 
{\hat r}_{n m}.\omega_1(n,m) f(x_{n,m}),\quad \omega_0(n,m)=i(\sigma_n\times\sigma_m)
\label{newomega0}
\eeq
\beq
\Omega_1=2g_V g_A\sum_{n,m}\tau_-(n)\tau_-(m) 
{\hat r}_{n m}\times\omega_1(n,m) f(x_{n,m}),\quad \omega_1(n,m)=\sigma_n-\sigma_m
\label{newomega1}
\eeq
 One can show that averaging over the initial polarizations and summing over the final ones one gets
 \beq
 |ME|^2=2 \sum_{i=0}^1|\prec f||\Omega_i^k||0^+\succ|^2,\quad k=i=J_f= \mbox{rank of the operator}
 \label{Eq:dbar}
 \eeq
 where the factor of 2 comes from the leptonic current, $k=J_f$ is the rank of the operator and the double bar indicates its standard reduced ME,
which can cause $ 0^+\rightarrow 1^-$ transitions.
 The mixing $V^2$ \cite{KSFF11} between the two atoms takes the form:
 \beq
 V^2=\Lambda^2_{\mbox{\tiny{eff}}} \left ( \prec {(0 s)^2}{a_B ^3}\succ \right )^2|ME|^2 
 \label{Eq:VV}
 \eeq
\section{Transitions to positive parity states in electron capture}
We will distinguish two cases:
\subsection{The captured electrons are in s states} 
As we have mentioned the $ 0^+\rightarrow 0^+$ transitions associated with double electron capture have already been studied during the last few years. The contribution arising from the neutrino mass independent mechanism has not been studied since the electron wave function cannot provide the extra parity change. This , however, can be accomplished via the recoil term in the nucleon current\footnote{It is interesting to note that in the case of neutrinoless double beta decay for the mass independent contribution one needs a correction either from the dipole term of the electron wave function or the nucleon recoil. Both are first order corrections. The nucleon recoil term has not been extensively studied even though it has been suggested by Kotani {\it al} long time ago \cite{KOTANI81}}. In other words one of the hadronic currents is of the form discussed above, but the other is of the form \cite{KOTANI81}:
 \beq
J^{\mu}_L(n)=\tau_-(n)(g_V {\bf D}_n,-g_A C_n),\quad J^{\nu}_R(m)=\tau_-(m)(g_V {\bf D}_m,+g_A C_m)
\eeq 
with
\beq
C_n= =\frac{1}{2 m_N}\sigma_n.(q_n-2P_n),\quad {\bf D}=\frac{1}{2 m_N}\left [(q_n-2P_n)=i \sigma_n \times q_n  \right ]
\eeq
Where $m_N$ and $P_n$ are the nucleon mass and momentum of nucleon $n$, while $q_n$ is the momentum transfer to that nucleon.

The procedure is a bit tedious but one can proceed as above  to arrive at
\begin{enumerate}
\item One operator associated with the leptonic current:
\beq
\vec{\ell}={\bar u}(p_1)\vec{\gamma} \gamma_5 u^c(p_2)
\eeq
which is associated with an operator appropriate for $0^+\rightarrow 1^+$ transitions, namely:
\beq
{\tilde \Omega}^{\mbox{\tiny{recoil}}}_1=\frac{G_F^2}{2 \sqrt{2}}\frac{i}{\pi R^2} 2 \sum_{n,m} \phi_{n_1,0}(r_n)\phi_{n_2,0}(r_m)\tau_-(n)\tau_-(m)  (\vec{\omega}_{nm}+\vec{\omega}'_{nm})
I(x(n,m))
\eeq
with
\beq
\vec{\omega}_{nm}= g_V g_A {\hat r}_{n m} (C_n-C_m) 
\eeq
\beq
\vec{\omega'}_{m n}=g_V^2 i ({\bf D}_n-{\bf D}_m)\times {\hat r}_{n,m}
\eeq
\item One operator associated with the leptonic current:
\beq
\ell_5={\bar u}(p_1) \gamma_5 u^c(p_2)
\eeq
which is suitable for  for $0^+\rightarrow 0^+$ transitions, namely:
\beq
{\tilde \Omega}^{\mbox{\tiny{recoil}}}_0=\frac{G_F^2}{2 \sqrt{2}}\frac{i}{\pi R^2_0}   2 \sum_{n,m}\phi_{n_1,0}(r_n)\phi_{n_2,0}(r_m)\tau_-(n)\tau_-(m)(\omega_{nm}+\omega'_{nm})
\eeq
\beq
\omega_{nm}=g_V g_A {\hat r}_{n m}.\left [ i( -\sigma_n\times D_m+\sigma_m\times D_n) \right]
\eeq
and
\beq
\omega'_{nm}=g_A^2 {\hat r}_{n m}.\left [ -\sigma_n C_m+\sigma_m C_n \right]
\eeq
\end{enumerate}
Under the factorization approximation one can write:
\beq
\tilde{\Omega}^{\mbox{\tiny{recoil}}}_i=\Lambda_{\mbox{\tiny{eff}}} \prec {(0 s)^2}{a_B ^3}\succ \Omega^{\mbox{\tiny{recoil}}}_i,\quad i=0,1
\eeq
\beq
 \Omega^{\mbox{\tiny{recoil}}}_1= 2 \sum_{n,m} \tau_-(n)\tau_-(m)  (\vec{\omega}_{nm}+\vec{\omega}'_{nm})
I(x(n,m)),
\eeq
\beq
 \Omega^{\mbox{\tiny{recoil}}}_0=  2 \sum_{n,m}\tau_-(n)\tau_-(m)(\omega_{nm}+\omega'_{nm}) I(x(n,m))
\eeq
 and then use Eqs (\ref{Eq:dbar}) and (\ref{Eq:VV}).

There remains the problem of transforming the operators $C_n$ and ${\bf D}_n$ into coordinate space. The operator ${\bf q}_n$ is local and leads to the operator $-i\frac{1}{R_0} \hat{r} \frac{d}{dx}$, where ${\bf x=r}/R_0,\,{\bf r}$ being the relative distance of the two nucleons. The operator $P_n$ is non local. Thus one finds:
\beq
P_n\rightarrow-i\frac{1}{R_0}\left (\overleftarrow {\bigtriangledown}_n- \overrightarrow {\bigtriangledown}_n\right)
\eeq
where the operator $\bigtriangledown_n$ operates with respect to ${\bf x}_n$ on the nuclear wave functions on the bra and ket as indicated by the arrows.
 As a simple example we will rewrite the operator the operator ${\hat r}_{nm}(C_n-C_m)$ in coordinate space. We find:
 \barr
 {\hat r}_{nm}(C_n-C_m)&=&\frac{-i}{2 M_n R_0}
  {\hat r}_{nm} \left \{ ({\hat r}_{nm}. \sigma_n-{\hat r}_{nm}. \sigma_m)\frac{d f}{dx}\right .\nonumber\\
  &-&\left . \left [ \left (\overleftarrow {\bigtriangledown}_nf(x)-f(x) \overrightarrow {\bigtriangledown}_n\right). \sigma_n-\left (\overleftarrow {\bigtriangledown}_mf(x)- f(x)\overrightarrow {\bigtriangledown}_m \right).\sigma_m  \right ] \right \}\nonumber\\
 \earr
 This can be rewritten by showing the tensor character of the various components as:
 \barr
 {\hat r}_{nm}(C_n-C_m)&=&\frac{-i}{2 M_n R_0}\left \{\left [-\frac{1}{\sqrt{3}}(\sigma_n-\sigma_m)-\frac{2 \sqrt{2}}{\sqrt{3}}\left (\left [{\hat r}_{nm}\times{\hat r}_{nm}\right]^2\otimes (\sigma_n-\sigma_m)\right)
 \right ]\frac{d f}{d x}\right .\nonumber\\
 &&\left .-\frac{1}{\sqrt{3}}\left ({\hat r}_{nm}.\left (\overleftarrow {\bigtriangledown}_nf(x)-f(x) \overrightarrow {\bigtriangledown}_n\right) \sigma_n-{\hat r}_{nm}.\left (\overleftarrow {\bigtriangledown}_mf(x)- f(x)\overrightarrow {\bigtriangledown}_m\right) \sigma_m\right ) \right . \nonumber\\
 &&\left .
 -\frac{2\sqrt{2}}{\sqrt{3}}\left (\left [{\hat r}_{nm}\otimes\left (\overleftarrow {\bigtriangledown}_n f(x)-f(x) \overrightarrow {\bigtriangledown}_n\right)\right ]^2\otimes \sigma_n \right. \right .\nonumber\\
 &-& \left. \left . \left [{\hat r}_{nm}\otimes\left (\overleftarrow {\bigtriangledown}_mf(x)- f(x)\overrightarrow {\bigtriangledown}_m\right)\right ]^2\otimes \sigma_m \right )
 \right \}
 \earr
In the case of the local  scalar spatial contribution  it is understood that:
\beq
\frac{d I}{d x}\rightarrow\frac{\delta(x)}{4 \pi x^2}
\eeq
 The overall tensor rank is $J=1$, the spin rank is one, but it contains both scalar and tensor components in the orbital sector. It can thus lead to $0^+\rightarrow 1^+$ transitions. The decomposition of the other operators, if needed,  can be done in a similar fashion. It is quite tedious but straightforward.
 
 The scale of the nucleon recoil contribution is set by $(1/2m_N R_0)=0.07 A^{-1/3}$. In the case of neurinoless double beta decay the square of this scale should be compared with the ratio of the phase space  of this $j_L \otimes j_R$ going through the electron momenta to that of the mass term.
 \subsection{One of the electrons is in a $p$ orbit.}
 If the resonance condition is met for such an electron, one may consider this possibility, even though the probability  for finding a $p$-electron inside the nucleus is tiny. The factorization approximation should not be used in this case and the electron wave function must be part of the nuclear transition operator. The average over the nucleus of the square of the electron wave function is, however, still a good measure of the scale of the expected nuclear matrix elements. Thus combining the above operator for the decays to negative parity states we find that one can have transitions to $0^+$, $1^+$ and $2^+$. The spin structure of the operator is still:
 \beq
  \omega_0=i \sigma_n\otimes\sigma_m \mbox{ or } \omega_1=\sigma_n-\sigma_m 
  \eeq
   i.e. antisymmetric of rank one. There appears to exist a wealth of orbital contributions.
 
  Indeed let us assume that the $p$ electron has total angular momentum $j$ and the initial angular momentum of the atom is $J$.
    The structure of effective transition operator can be cast in the form:
  \beq
  \omega_0.{\hat r} \ell_0 \rightarrow U(1,1/2,j,0,1/2,1/2,1,0,J)\delta_{J,1} \left [\omega_0.{\hat r}  \vec{V}\right] \ell_0
  \eeq
  for the time component and
   \barr
  (\omega_1.\times {\hat r}).\vec{\ell} \rightarrow&&U(1,1/2,j,0,1/2,1/2,1,0,J)U(1,1,0,1,1,J,J,0,J) \nonumber\\
  &&\left [ (\omega_1\times {\hat r}) \otimes \vec{V}\right]^J \vec{\ell}.\vec{s}
  \earr
  for the space component ($\vec{\ell}$ is the leptonic current not to be confused with the orbital angular momentum operator). $ U(...)$ are the usual unitary nine-j symbols needed for the re-coupling of the angular momenta involved and $\vec{V}$ essentially is the two electron orbital wave function:
  \beq
  \vec{V}=\frac{1}{\sqrt{4 \pi}}\frac{\sqrt{3}}{\sqrt{4 \pi}}\frac{1}{\sqrt{2}} \left (A^{''} {\bf x}_2+A^{'} {\bf x}_1\right )
  \eeq
  with
  \beq
  A^{''}= \frac{R_0}{r_m}\left (\phi_{n_1,0}(r_n),\phi_{n_2,1}(r_m)\right ),\quad A^{'}=\frac{R_0}{r_n}\left (\phi_{n_1,1}(r_n),\phi_{n_2,0}(r_m)\right )
  \eeq
  with $\phi_{n_i,\ell_i} $, $i=1,2$  the radial electron wave function. We have chosen to express the length in units of the nuclear radius $R_0$, i.e. ${\bf x}={\bf r}/R_0$, with compensating factors in the definition of $A^{'}$ and $A^{''}$, a procedure quite standard in the evaluation of the nuclear ME. 
  
  We found it convenient to go into the relative ${\bf x}={\bf x_n}-{\bf x_m}$ and center of mass $2{\bf X}={\bf x_n}+{\bf x_m}$ coordinates and thus 
   we get:
  \beq
  \vec{V}=\frac{1}{4 \pi}\left (A(n,m) {\bf x}+B(n,m) {\bf X}\right ),\quad
    A(n,m)=\sqrt{3}\frac{1}{\sqrt{2}} \frac{(A^{'}-A^{''})}{2},
     \quad B(n,m)= \sqrt{3}\frac{1}{\sqrt{2}}  (A^{'}+A^{''})
  \eeq
  The operator $ \vec{V}$ must be symmetric under the interchange of particles $n$ and $m$ to yield a total  operator which is overall symmetric under the exchange of the particles $n$ and $m$. So only the B-terms survive.
  
  The next step is to perform an additional re-coupling bringing together the two spacial parts. One then finds:
  \begin{enumerate}
  \item For the time component:
  \beq
  \omega_0.{\hat r} \ell_0 \rightarrow \frac{B(n,m)}{4 \pi} \sum_{L}C_0(j,L) [\omega_0\otimes T^{L}(\hat{r},\vec{R})]^J\,\delta_{J,1}\ell_0
  \label{full0}
  \eeq
  with 
  $$
  C_0(3/2,L)=-\frac{1}{\sqrt{2}}C_0(1/2,L)=(\frac{\sqrt{2}}{3\sqrt{3}},-\frac{1}{\sqrt{6}},-\frac{\sqrt{2}}{\sqrt{15}}\mbox{ for } L=0,1,2
  $$
  and
  \beq
  T^0={\hat r}.{\bf X},\quad T^1= {\hat r}\,\times {\bf X},\quad  T^{2}= [{\hat r}\otimes {\bf X}]^2
  \eeq
  \item For the space component:
  \beq
  \omega_1\times{\hat r} \ell_0 \rightarrow \frac{B(n,m)}{4 \pi} \sum_{L}C_2(j,J,L) [\omega_1\otimes T^{L}(\hat{r},\vec{R})]^J\vec{\ell}.\vec{s}
    \label{full1}
    \eeq
  where the coefficients $C_2(j,J,L)$ are given in table \ref{tabC2}
  \end{enumerate}
  By combining Eqs (\ref{omega0}) and (\ref{full0}) as well as Eqs (\ref{omega1}) and (\ref{full1}) we obtain the complete operators relevant for the decay of $0^+$ to positive parity states with $J=0,1$ and 2.
  
  To simplify matters  one may seek an approximation familiar from the $0s$ case, i.e. to take the radial part of the electron wave function outside of the integrals and replace it with its average value over the nucleus. Thus:
  \beq
\tilde{\Omega}_i=\Lambda_{\mbox{\tiny{eff}}} \sqrt{\prec a_B^6 {B^2}\succ}\Omega_i, \quad \Lambda_{\mbox{\tiny{eff}}}=\quad \frac{G_F^2}{2 \sqrt{2}}\frac{1}{\pi R^2_0 a_B^3},\quad B\leftrightarrow B(n,m)
\eeq
where $a_B$ is Bohr radius and, again, the averaging is over the radial part alone. Thus the effective transition operators now become:
\beq
\Omega_0=4 g^2_A\sum_{n,m}\tau_-(n)\tau_-(m)  f(x_{n,m})\sum_{L}C_0(j,L) [\omega_0\otimes T^{L}(\hat{r},\hat{R})]^J\,\delta_{J,1}
\label{plusomega0}
\eeq
\beq
\Omega_1=2g_V g_A\sum_{n,m}\tau_-(n)\tau_-(m)  f(x_{n,m})
\sum_{L}C_2(j,J,L) [\omega_1\otimes T^{L}(\hat{r},\hat{R})]^J
\label{plusomega1}
\eeq
In fact it will be adequate to consider
$$
\prec B^2\succ \longrightarrow\prec \left (R_{n,0}\right)^2\,\succ  \prec 3 \left (\frac{R_0}{r}R_{n,1}\right)^2\succ
$$
where $R_{n,\ell}$ are the radial parts of the bound electron wave functions. Note that the $p$ wave function involved in the averaging has been stripped of an extra power of $r$, since the $\vec{r}$ has been incorporated into the nuclear matrix element.

 The expression for the nuclear ME is the same with that of Eq. (\ref{Eq:dbar}), where $k=J=J_f$ is the rank of the operator and the coupling of the two atoms is given by:
 \beq
 V^2=\Lambda^2_{\mbox{\tiny{eff}}}\prec a_B^6 {B^2}\succ|ME|^2
 \eeq

\begin{figure}
\begin{center}
\subfloat[]
{
\rotatebox{90}{\hspace{0.0cm} Normalized W-S:$\rho\rightarrow$}
\includegraphics[height=.15\textheight]{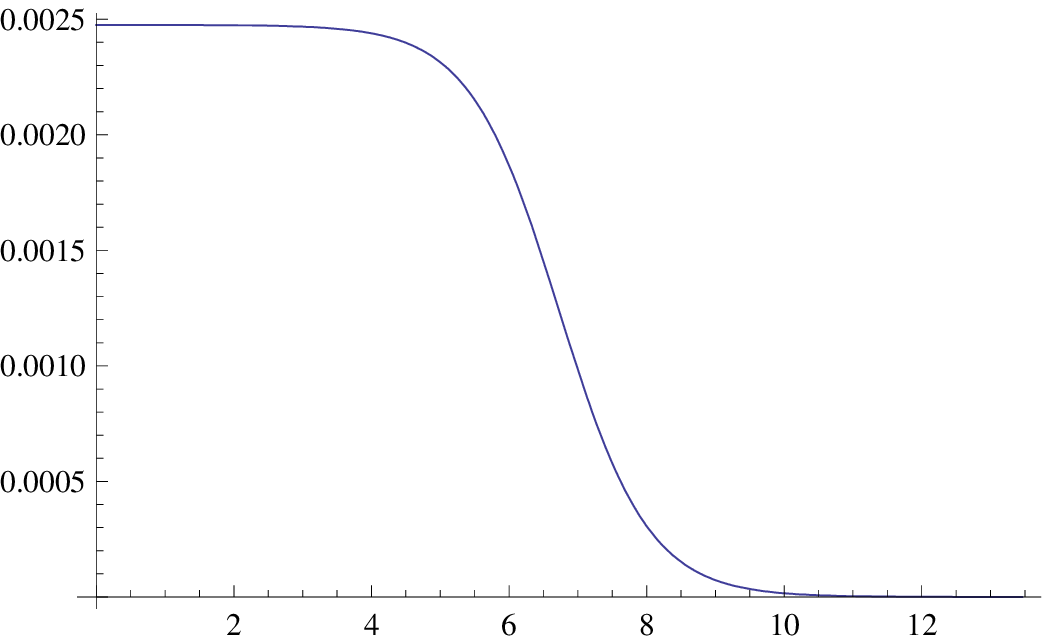}
}
\subfloat[]
{
\rotatebox{90}{\hspace{0.0cm}{ Normalized}: W-S :$r^2 \rho\rightarrow$}
\includegraphics[height=.15\textheight]{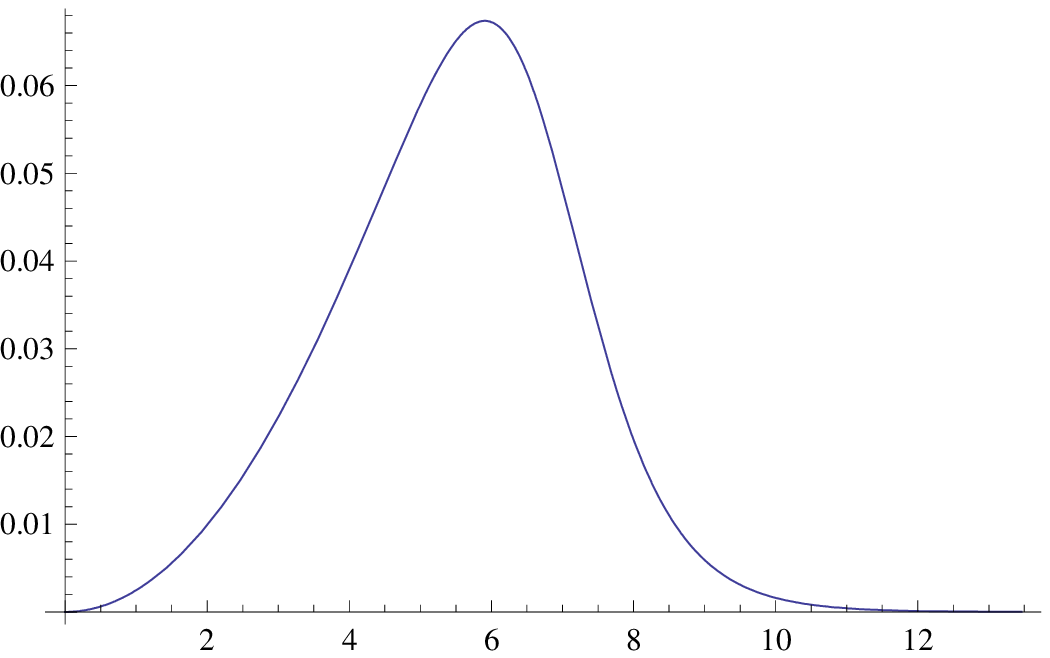}
}\\
{\hspace{-2.0cm} $r\rightarrow$fm}
\caption{ The (normalized) Woods-Saxon density profile employed in obtaining the average electron wave functions.
 \label{fig:nucden}}
\end{center}
\end{figure}

\begin{figure}
\begin{center}
\subfloat[]
{
\rotatebox{90}{\hspace{0.5cm} $\Psi_{SM}(r)^2\rightarrow$}
\includegraphics[height=.15\textheight]{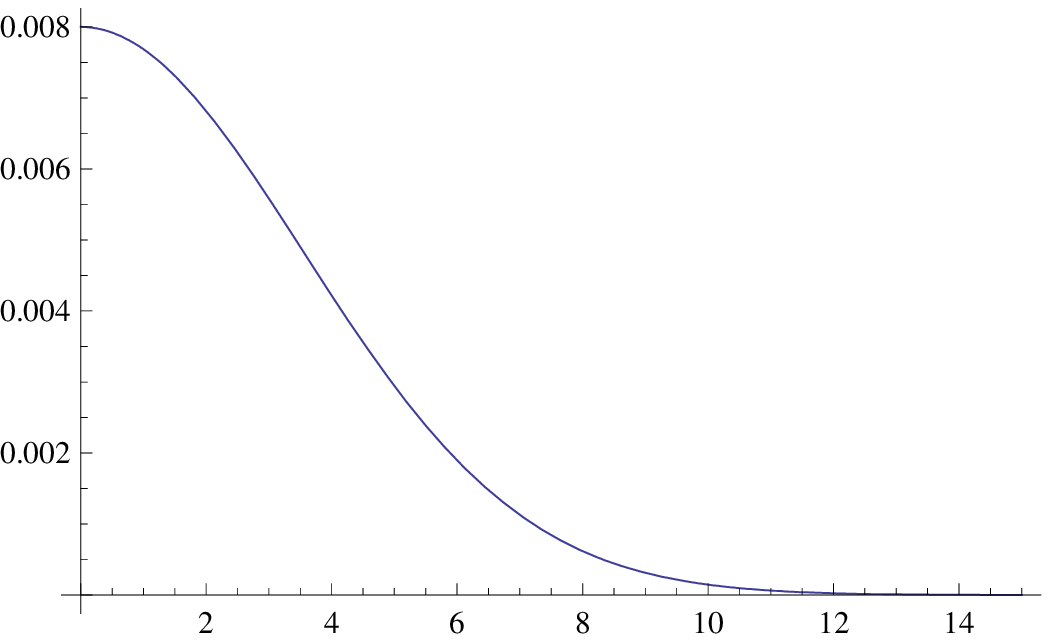}
}
\subfloat[]
{
\rotatebox{90}{\hspace{0.5cm} $r^2 \Psi_{SM}(r)^2\rightarrow$}
\includegraphics[height=.15\textheight]{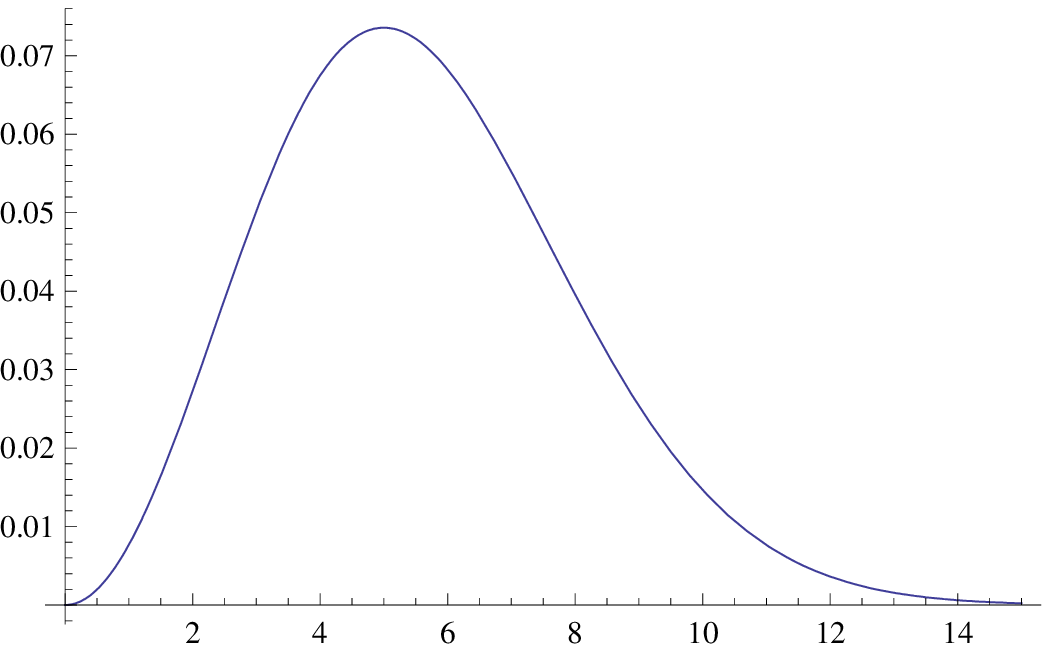}
}\\
{\hspace{-1.0cm} $r\rightarrow$fm}
\caption{ The nuclear S.M. density profile employed in in obtaining the average electron wave functions and checking the factorization approximation.
 \label{fig:SMden}}
\end{center}
\end{figure}

 We will check the factorization approximation using realistic atomic $3s$ and $4p_{3/2}$  wave functions provided to us by  Shabaev
  \cite{Shabaev}, which were obtained by standard methods (see, e.g., Grant \cite{GRANT70} and Bratsev {\it et al} \cite{Bratsev77}). 
\begin{itemize}
\item Using the density profile shown in Fig. \ref{fig:nucden}, corresponding  to a phenomenological normalized to unity Woods-Saxon density,  we  find:
$$
\prec a_B^3 \left ( R_{3s}\right)^2\succ=2.4\times10^{3},\quad\prec 3 a_B^3 \left ( (\frac{R_0}{r}R_{4p_{3/2}} \right )^2\succ=3.2\times10^{-2}\longrightarrow$$
$$ \prec 3 \left (\frac{R_0}{r} R_{4p_{3/2}} \right )^2\succ/\prec \left ( R_{3s}\right)^2\succ=1.3\times 10^{-5}
$$
\item Using the density profile corresponding to  a simplified S.M. wave  function shown in Fig. \ref{fig:SMden}, we find:
$$
\prec a_B^3 \left ( R_{3s}\right)^2\succ=1.4\times10^{4},\quad\prec 3 a_B^3 \left (\frac{R_0}{r} R_{4p_{3/2}} \right )^2\succ=1.8\times10^{-1}\longrightarrow $$
$$\prec3\left ( (\frac{R_0}{r}R_{4p_{3/2}} \right )^2\succ/\prec \left ( R_{3s}\right)^2\succ=1.4\times 10^{-5},
$$
\end{itemize}
Had we not used the factor $3 (R_0/r)^2$ the $p$-integrals would have been quite a bit smaller.
The agreement in the ratio of $4p$ to $3s$ probabilities between the two nuclear densities is quite good. Furthermore, using a simple $0s$ HO wave function, we find that the factorization approximation works to within $10\%$. More detailed computations using realistic and more accurate shell model wave functions are currently underway. The results, with and without employng the factorization approximation, will be published elsewhere \cite{CivSuhVer11}. The nuclear matrix elements leading to the $^{156}$Dy$\rightarrow^{156}$Gd leading to sharp resonaces \cite{EliNov} $0^-,1^-,0^+,1^+$ and $2^+$ will also be calculated \cite{CivSuhVer11}, using realistic wave functions both for the intial and the final nuclear states.

  \begin{table}
  \caption{The coefficients $C_2(j,J,L)$ discussed in the text}
  \label{tabC2}
  \begin{tabular}{|c|c|c|c|}\hline\hline
  &&&\\
  j&J&L&$C_2(j,J,L)$\\
   \hline
   $\frac{1}{2}$ & 0 & 1 & $-\frac{1}{2}$ \\
 $\frac{1}{2}$ & 1 & 0 & $-\frac{\sqrt{2}}{9}$ \\
 $\frac{1}{2}$ & 1 & 1 &$ -\frac{1}{2 \sqrt{6}}$\\
 $\frac{1}{2}$ & 1 & 2 & $\frac{\sqrt{\frac{5}{2}}}{3}$ \\
 $\frac{3}{2}$ & 1 & 0 & $-\frac{1}{9}$ \\
 $\frac{3}{2}$ & 1 & 1 & $-\frac{1}{4 \sqrt{3}}$ \\
$ \frac{3}{2}$ & 1 & 2 & $\frac{\sqrt{5}}{6}$\\
$ \frac{3}{2}$ & 2 & 1 & $\frac{1}{4}$ \\
 $\frac{3}{2}$ & 2 & 2 & $\frac{\sqrt{3}}{2}$\\
 \hline
  \hline
  \end{tabular}
  \end{table} 
 \section{Discussion}
  As we have mentioned the observation of neutrinoless double decay will be a great event in particle and nuclear physics. It will establish the existence of flavor violating interactions and it will demonstrate that, independently of the dominant mechanism causing it, the neutrinos are Majorana particles. More importantly, it is expected to  complement the great successes on neutrino oscillations by determining the elusive up to now absolute scale of neutrino mass. For this last great contribution to become unambiguous, it will be necessary to determine what fraction of this decay  is due to the light neutrino mass compared to  all the other competing mechanisms \cite{JDV02}. That is  a very difficult task indeed. It may become feasible, if there exist data in as many targets as possible. In fact it has recently been shown that, within the mass terms alone, even if both  left and right handed currents are involved, this is, in principle, possible \cite{SVF10,FMPSV11}. Provided, of course, that experimental data involving a sufficient number of targets become available. It will be much more difficult to achieve this, if the mass independent contribution involving the $j_L\times j_R$ contribution becomes important. In this direction the observation of double electron capture will be very useful. More specifically 
 we have seen that:
 \begin{enumerate}
 \item The neutrino mass term can proceed in neutrinoless double electron capture as in ordinary double beta decay. For the latter to be practical, of course, the resonance condition must be met.
 \item Neutrinoless transitions to $0^+\rightarrow J^{\pm}\ne 0^+$ can occur only if there exist right handed interactions through the right left interference in the leptonic sector $(j_L \otimes j_R)$. Then:
 \begin{itemize}
 \item The unusual transitions to the negative parity $0^-$ and $1^-$ sates can naturally occur  in 
 double electron capture,
 since they are energetically allowed, if the resonance condition is met. So such an observation will unambiguously determine the neutrino mass independent lepton violating parameter, which is hard to get in neutrinoless double beta decay.
 \item Transitions to positive parity states $\ne 0^{+}$ in ordinary double beta decay
   can occur either through the dipole component of the electron wave functions, the most favored scenario, or if one of the hadronic currents involved  contains the parity changing nucleon recoil terms.
 \item Transitions to positive parity states with angular momentum less than 3 in neutrinoless double electron capture can proceed if a)  one of the hadronic currents involves  the recoil term, while both   electrons are  in s-states or b) more naturally via the negative parity operator involving  ordinary nucleon currents, 
 if one  s-electron and one p-electron are involved in the capture.
  The probability of finding a $p$-electron in the nucleus is, however, quite small, but this disadvantage maybe overcome by the collaborative effect of the expected large nuclear matrix elements and the resonance condition for capture to $2^{+}$.
 \item The transition operators involved have a reasonable structure, but nuclear matrix elements of such operators have not yet been calculated for any system. We have no reason to expect them to be suppressed. The nuclear structure calculations involving such nuclear systems may be a true challenge to nuclear theorists. 
 \end{itemize}
 \end{enumerate}
 It is clear from the above that the neutrinoless double electron capture can distinguish between the mass term contribution and the neutrino mass independent $j_L\times j_R$ interference term. It can do it more efficiently than ordinary beta decay due to the possibility of the negative parity nuclear transitions.
 It cannot, of course, distinguish between the various mass terms (light neutrino, heavy neutrino, SUSY contribution etc), but neither can ordinary double beta decay. This can only  be done by exploiting the results of at least four different experiments (leading to $0^+$), or a suitable combination of a sufficient number of both ground state and exited $0^+$ state transitions. In this multiple experiment analysis, double electron capture can be come a very useful partner.
	\section*{Acknowledgments} The author is  is indebted Professor Shabaev for providing the needed atomic wave functions and to Dr S. Eliseev and Professor Y. Novikov for discussions and useful comments. The hospitality of Professor K. Blaum in the Kernphysik, MPI-Hd and of the Theory Division at CERN are happily acknowledged.
 	\section*{Bibliography}

\end{document}